\documentclass{article}
\usepackage{hyperref}

\begin{document}

\begin{center}

{\large {\bf  Proceedings of the II Iberian Nuclear Astrophysics Meeting on Compact Stars},\\ September 22$^{nd}$-23$^{rd}$ 2011, Salamanca, Spain. \\}

\end{center}
Editors: M. A. P\'erez-Garc\'\i a, J. A. Pons,  C. Albertus,Journal of Physics: Conference Series, Volume 342 (2012), ISSN 1742-6588 .\\

{\bf ORGANIZING COMMITTEE\\}

\noindent Dr M \'Angeles P\'erez-Garc\'\i a (\'Area F\'\i sica Te\'orica-Universidad de Salamanca \& IUFFYM)\\
Dr J A Miralles (Universidad de Alicante)\\
Dr J Pons (Universidad de Alicante)\\
Dr C Albertus (\'Area F\'\i sica Te\'orica-Universidad de Salamanca \& IUFFYM)\\
Dr F Atrio (\'Area F\'\i sica Te\'orica-Universidad de Salamanca \& IUFFYM)\\\\

{\bf PREFACE\\}

The second Iberian Nuclear Astrophysics meeting was held at the
University of Salamanca, Spain on September 22$^{nd}$-23$^{rd}$ 2011. This volume index contains the links to most of the presentations delivered at this international workshop. This meeting was the second in the series following the
previous I Encuentro Ib\'erico de Compstar, held at the University of
Coimbra, Portugal in 2010.

The main purpose of this meeting was to strengthen the scientific
collaboration between the participants of the Iberian and the rest of
the southern European branches of the European Nuclear Astrophysics
network, formerly, COMPSTAR. This ESF (European Science Foundation)
supported network has been crucial in helping to make a broader
audience for the the most interesting and relevant research lines
being developed currently in Nuclear Astrophysics, especially related
to the physics of neutron stars. It is indeed important to emphasize
the need for a collaborative approach to the rest of the scientific
communities so that we can reach possible new members in this
interdisciplinary area and as outreach for the general public.

The program of the meeting was tailored to theoretical descriptions of
the physics of neutron stars although some input from experimental
observers and other condensed matter and optics areas of interest was
also included. The main scientific topics included:

   \begin{itemize}
    \item{Magnetic fields in compact stars}
 \item{    Nuclear structure and in-medium effects in nuclear interaction}
 \item{    Equation of state: from nuclear matter to quarks}
 \item{    Importance of crust in the evolution of neutron stars}
 \item{    Computational simulations of collapsing dense objects}
 \item{    Observational phenomenology}
   \end{itemize}

In particular, leading experts from the computational simulation of
core-collapse supernovae and the effect of hadron–quark phase
transitions developed specialized review talks. Prospects in future
observations or a more dilute classification of magnetars were also
discussed. The importance of the equation of state, three-body forces,
finite nuclei, phenomenological fermionic interaction models, and the
microphysics inputs of different many-body approaches to some very
important quantities as the symmetry energy were reviewed and
discussed from either the non-relativistic to the relativistic
framework. The importance of the crust with the existence of a
crystallized structure and vortex-crust pinning were some of the
important subjects discussed in the context of cooling and field
dynamics.

Finally, some condensed matter and optics talks presented us the rich
insight that Cold Atom Physics can give us on low-density interactions
and the new and very intense laser Petawatt beams can test matter
under strong external fields, respectively.

We would to thank the Faculty of Science and University of Salamanca
for hosting the meeting. We also thank for partial financial support
the European ERC Network COMPSTAR, {\it The Physics of Neutron Stars} under
reference 3803 and the Spanish Ministerio de Ciencia e Innovaci\'on
(MICINN) with project FIS2011-14759, MULTIDARK Consolider-Ingenio 2010, MICINN ref. CSD2009-00064 and the local institutions of
Instituto de F\'isica Fundamental y Matem\'aticas (IUFFYM) and Universidad
de Salamanca, Spain.

 Of course we thank those who have contributed to
make this meeting a nice occasion to gather and start to develop
fruitful collaborations. To them go our grateful acknowledgments.\\\\

{\bf INDEX}

\begin{itemize}

\item Evolution of proto-neutron stars with hadron–quark phase transition. \\
 I Bombaci, D Logoteta, C Provid\^encia and I Vida\~na. \\
 \href{http://dx.doi.org/10.1088/1742-6596/342/1/012001}{doi:10.1088/1742-6596/342/1/012001}

\item The Symmetry energy of nuclear matter under a strong magnetic field.\\
 R Casali, C Provid\^encia and D Menezes.\\
 \href{http://dx.doi.org/10.1088/1742-6596/342/1/012002}{doi:10.1088/1742-6596/342/1/012002}

\item Unified equation of state for neutron stars and supernova cores
  using the nuclear energy-density functional theory. \\
 A F Fantina, N Chamel, J M Pearson and S Goriely. \\
  \href{http://dx.doi.org/10.1088/1742-6596/342/1/012003}{doi:10.1088/1742-6596/342/1/012003}

\item Vortex-lattice interaction in Pulsar Glitches.\\
 F Grill and P Pizzochero. \\
\href{http://dx.doi.org/10.1088/1742-6596/342/1/012004}{doi:10.1088/1742-6596/342/1/012004}

\item Structure and Shear Modulus of the Neutron Star Crust.\\
 J Hughto.\\
 \href{http://dx.doi.org/10.1088/1742-6596/342/1/012005}{doi:10.1088/1742-6596/342/1/012005}

\item Effect of hyperonic three-body forces on the maximum mass of
  neutron stars. \\ 
 D Logoteta, I Vida\~na, C Provid\^encia, A Polls and I Bombaci. \\
 \href{http://dx.doi.org/10.1088/1742-6596/342/1/012006}{doi:10.1088/1742-6596/342/1/012006}

\item Is the apparent dichotomy between bursting activity of magnetars
  and radio pulsars real?\\ 
  J A Pons and R Perna. \\
  \href{http://dx.doi.org/10.1088/1742-6596/342/1/012007}{doi:10.1088/1742-6596/342/1/012007}

\item Effect of the symmetry energy on compact stars.\\
 C Providência, R Cavagnoli, D P Menezes and P K Panda.\\
 \href{http://dx.doi.org/10.1088/1742-6596/342/1/012008}{doi:10.1088/1742-6596/342/1/012008}

\item The pygmy dipole strength, the neutron radius of 208Pb and the
  symmetry energy.\\ 
  X Roca-Maza, M Brenna, M Centelles, G Col\`o, K Mizuyama, G Pozzi, X Vi\~nas and M Warda.\\ 
  \href{http://dx.doi.org/10.1088/1742-6596/342/1/012009}{doi:10.1088/1742-6596/342/1/012009}

\item Intense infrared lasers and laboratory astrophysics. \\
 L Roso.\\
 \href{http://dx.doi.org/10.1088/1742-6596/342/1/012010}{doi:10.1088/1742-6596/342/1/012010}

\item Fully antisymmetrised dynamics for bulk fermion systems.\\
 K Vantournhout and H Feldmeier.\\
 \href{http://dx.doi.org/10.1088/1742-6596/342/1/012011}{doi:10.1088/1742-6596/342/1/012011}

\item Symmetry energy within the BHF approach. \\
 I Vida\~na, C Provid\^encia, A Polls and A Rios. \\
 \href{http://dx.doi.org/10.1088/1742-6596/342/1/012012}{doi:10.1088/1742-6596/342/1/012012}

\item The influence of magnetic field geometry on magnetars X-ray spectra. \\
 D Vigan\`o, N Parkins, S Zane, R Turolla, J A Pons and J A Miralles.\\
 \href{http://dx.doi.org/10.1088/1742-6596/342/1/012013}{doi:10.1088/1742-6596/342/1/012013}

\end{itemize}

\end{document}